\documentclass{article}\usepackage{aaecc99} 

\usepackage{rotate}
\usepackage{amsfonts}

\def\ket#1{|#1\rangle}
\def\DFT{{\rm DFT}}

\def\F{\mathbb{F}}
\def\C{\mathbb{C}}
\def\binexp{{{\cal B}}}
\def\dualbinexp{{{\cal B}^\bot}}

\def\bm#1{\mathchoice{\mbox{\boldmath{$\displaystyle #1$}}}%
{\mbox{\boldmath{$\textstyle #1$}}}%
{\mbox{\boldmath{$\scriptstyle #1$}}}%
{\mbox{\boldmath{$\scriptscriptstyle #1$}}}}

\newtheorem{theorem}{Theorem}
\newtheorem{definition}[theorem]{Definition}

\begin{document}
% \aaeccAfour             % : uncomment for A4 paper

\aaecctitle       {Quantum Reed--Solomon Codes}

\aaeccauthor     {\ \hspace{-4cm}Markus Grassl, Willi Geiselmann, and
Thomas Beth\hspace{-4cm}\ } 
{\centerline{Institut f{\"u}r Algorithmen und Kognitive Systeme}
                  Universit{\"a}t Karlsruhe\\ 
                  76\,128 Karlsruhe, Germany\\
\centerline{Email: $\{$grassl, geiselma, EISS\_Office$\}$@ira.uka.de}}
\aaeccmaketitle

\begin{aaeccabstract}
We introduce a new class of quantum error--correcting codes derived
from (classical) Reed--Solomon codes over finite fields of
characteristic two. Quantum circuits for encoding and decoding based
on the discrete cyclic Fourier transform over finite fields are
presented.
\end{aaeccabstract}

\begin{aaeccpaper}
\aaeccsection{Introduction}
\smallskip
During the last years it has been shown that computers taking
advantage of quantum mechanical phenomena outperform currently used
computers. The striking examples are integer factoring in polynomial
time (see~\cite{Shor94}) and finding pre--images of an $n$--ary
Boolean function (``searching'') in time $O(\sqrt{2^n})$
(see~\cite{Gro96_search}). Quantum computers are not only of
theoretical nature---there are several suggestions how to physically
realize them (see, e.\,g., \cite{CiZo95,CFH96}).

On the way towards building a quantum computer, one very important
problem is to stabilize quantum mechanical systems since they are very
vulnerable. A theory of quantum error--correcting codes has already
been established (see~\cite{KnLa97}). Nevertheless, the problem of how
to encode and decode quantum error--correcting codes has hardly been
addressed, yet.

We present the construction of quantum error--correcting codes based
on classical Reed--Solomon (RS) codes. For RS codes, many classical
decoding techniques exist. RS codes can also be used in the context of
erasures and for concatenated codes. Encoding and decoding of quantum
RS codes is based on quantum circuits for the cyclic discrete Fourier
transform over finite fields which are presented in the full paper,
together with the quantum implementation of any linear transformation
over finite fields. We start with a brief introduction to quantum
computation and quantum error--correcting codes, followed by some
results about binary codes obtained from codes over extension fields.

\smallskip
\aaeccsection{Qubits and Quantum Registers}\label{qubits}
\smallskip
The basic unit of quantum information, a {\em quantum bit} (or short
{\em qubit}), is represented by the normalized linear combination 
\begin{equation}\label{qubit}
\ket{q}=\alpha\ket{0}+\beta\ket{1},\qquad\mbox{where
$\alpha,\beta\in\C$, $|\alpha|^2+|\beta|^2=1$.}
\end{equation}
Here $\ket{0}$ and $\ket{1}$ are orthonormal basis states written in
Dirac notation. The normalization condition in Eq.~(\ref{qubit}) stems
from the fact that when extracting classical information from the
quantum system by a measurement, the results ``0'' and ``1'' occur
with probability $|\alpha|^2$ and $|\beta|^2$, resp.

A {\em quantum register} of length $n$ is obtained by combining $n$
qubits modelled by the $n$--fold tensor product $(\C^{\,2})^{\otimes
n}$. The canonical orthonormal basis of $(\C^{\,2})^{\otimes n}$ is
$$
B:=\bigl\{
\ket{b_1}\otimes\ldots\otimes\ket{b_n}=:\ket{b_1\ldots
b_n}=\ket{\bm{b}}\bigm|b_i\in\{0,1\}\bigr\}.
$$
Hence the state of an $n$ qubit register is given by
$$
\ket{\psi}=\sum_{\bm{b}\in\{0,1\}^n}c_{\bm{b}}\ket{\bm{b}},\qquad
\mbox{where $c_{\bm{b}}\in\C$ and $\sum\limits_{\bm{b}\in\{0,1\}^n}
|c_{\bm{b}}|^2=1.$}
$$
\smallskip
\aaeccsection{Quantum Error--Correcting Codes}
\smallskip
One common assumption in the theory of quantum error--correcting codes
is that errors are local, i.\,e., only a small number of qubits are
disturbed when transmitting or storing the state of an $n$ qubit
register. The basic types of errors are bit--flip errors exchanging
the states $\ket{0}$ and $\ket{1}$, phase--flip errors changing the
relative phase of $\ket{0}$ and $\ket{1}$ by $\pi$, and their
combination. The bit--flip error corresponds to the Pauli matrix
$\sigma_x$, the phase--flip error to $\sigma_z$, and their combination
to $\sigma_y$. It is sufficient to consider only this discrete set of
errors in order to cope with any possible local error (see
\cite{KnLa97}).

Errors operating on an $n$ qubit system are represented by tensor
products of Pauli matrices and identity. The {\em weight\/} of an
error $\bm{e}=e_1\otimes \ldots \otimes e_n$, where
$e_i\in\{id,\sigma_x, \sigma_y,\sigma_z\}$ is the number of local
errors $e_i$ that differ from identity.

The construction of quantum Reed--Solomon codes is based on the
construction of quantum error--correcting codes from weakly self--dual
binary codes (see, e.\,g., \cite{Ste96:multiple}). That construction
is summarized by the following definition and theorem.
\begin{definition}\label{qcode}
Let $C=[N,K]$ be a weakly self--dual linear binary code, i.\,e., $C\le
C^\bot$, and let $\{\bm{w}_j\mid j=1,\ldots, 2^{N-2K}\}$ be a system
of representatives of the cosets $C^\bot/C$. Then the basis states
of a quantum code ${\cal C}=[[N,N-2K]]$ are given by
$$
\ket{\psi_j}=\frac{1}{\sqrt{|C|}}
  \sum_{\bm{c}\in{C}}\ket{\bm{c}+\bm{w}_j}.
$$
\end{definition}
\begin{theorem}\label{qmindist}
Let $d$ be the minimum distance of the dual code $C^\bot$ in
Definition~\ref{qcode}. Then the corresponding quantum code is capable
of detecting up to $d-1$ errors or, equivalently, is capable of
correcting up to $(d-1)/2$ errors.
\end{theorem}
\smallskip
\aaeccsection{Main Results}
\smallskip
The following definition and theorem show how to obtain 
weakly self--dual binary codes from codes over extension fields.
\begin{definition} Let ${C}=[N,K,D]$ denote a linear code of length
$N$, dimension $K$, and minimum distance $D$ over the field
$\F_{2^k}$, and let ${\cal B}=(b_1,\ldots,b_k)$ be a basis of
$\F_{2^k}$ over $\F_2$. Then the {\em binary expansion} of ${C}$ with
respect to the basis ${\cal B}$, denoted by $\binexp(C)$, is the
linear binary code $C_2=[kN,kK,d\ge D]$ given by
$$
C_2=\binexp(C):=
\bigl\{ \left(c_{ij}\right)_{i,j}\in\F_2^{\,kN} \bigm|
  \bm{c}=\left(\textstyle\sum_j c_{ij} b_j\right)_i\in{C}\bigr\}.
$$
\end{definition}
\begin{theorem}
Let ${C}=[N,K]$ be a linear code over the field $\F_{2^k}$ and let
${C}^\bot$ be its dual. Then the dual code of the binary expansion
$\binexp({C})$ of ${C}$ with respect to the basis ${\cal B}$ is the
binary expansion $\dualbinexp({C}^\bot)$ of the dual code $C^\bot$
with respect to the dual basis ${\cal B}^\bot$, i.\,e., the following
diagram is commutative:
$$\def\arraystretch{1.5}
\begin{array}{rccll}
&{C} & \longrightarrow & {C}^\bot\\
\mbox{\footnotesize basis ${\cal B}$}
 &\left\downarrow\rule{0pt}{10pt}\right. && 
  \left\downarrow\rule{0pt}{10pt}\right. 
 &\mbox{\footnotesize dual basis ${\cal B}^\bot$}\\
&\llap{$\binexp{}$}(C) & \longrightarrow & 
   \dualbinexp\rlap{$({C}^\bot)=\binexp({C})^\bot$}\\
\end{array}
$$
\end{theorem}
Using these results, we are ready to define quantum Reed--Solomon
codes, based on classical weakly self--dual RS codes.
\begin{definition}\label{QRScode}
Let $C=[N,K,\delta]$ where $N=2^k-1$, $K=N-\delta+1$, and
$\delta>N/2+1$ be a Reed--Solomon code over $\F_{2^k}$ (with
$b=0$). Furthermore, let ${\cal B}$ be a self--dual basis of
$\F_{2^k}$ over $\F_2$. Then the {\em quantum Reed--Solomon code} is
the quantum error--correcting code ${\cal C}$ of length $kN$ derived
from the weakly self--dual binary code $\binexp(C)$ according to
Definition~\ref{qcode}.
\end{definition}
The parameters of the quantum Reed--Solomon code are given by the following
theorem. 
\begin{theorem}\label{QRSpara}
The quantum RS code ${\cal C}$ of Definition~\ref{QRScode} encodes
$k(N-2K)$ qubits using $kN$ qubits. It is able to detect at least up
to $K$ errors, i.\,e., ${\cal C}=[[kN,k(N-2K),d\ge K+1]]$.
\end{theorem}

\begin{figure}[hbt]
\centerline{\unitlength0.9pt\scriptsize
\begin{picture}(30,115)(-20,0)
\multiput(0,0)(0,20){2}{\line(1,0){10}}
\multiput(0,30)(0,25){2}{\line(1,0){10}}
\multiput(0,65)(0,20){2}{\line(1,0){10}}
\multiput(0,95)(0,15){2}{\line(1,0){10}}
\multiput(5,6)(0,4){3}{\makebox(0,0){.}}
\multiput(5,38.5)(0,4){3}{\makebox(0,0){.}}
\multiput(5,71)(0,4){3}{\makebox(0,0){.}}
\multiput(5,100)(0,2.5){3}{\makebox(0,0){.}}
\put(-5,0){\makebox(0,0)[r]{$\ket{0}$}}
\put(-5,20){\makebox(0,0)[r]{$\ket{0}$}}
\put(-5,65){\makebox(0,0)[r]{$\ket{0}$}}
\put(-5,85){\makebox(0,0)[r]{$\ket{0}$}}
\put(-5,110){\makebox(0,0)[r]{$\ket{\phi_1}$}}
\put(-5,95){\makebox(0,0)[r]{$\ket{\phi_k}$}}
\put(-5,55){\makebox(0,0)[r]{$\ket{\phi_{k+1}}$}}
\put(-5,30){\makebox(0,0)[r]{$\ket{\phi_{k(N-2K)}}$}}
\end{picture}%
\begin{picture}(20,100)
\multiput(0,30)(0,25){2}{\line(1,0){20}}
\multiput(0,65)(0,20){2}{\line(1,0){20}}
\multiput(0,95)(0,15){2}{\line(1,0){20}}
\multiput(0,0)(0,20){2}{\line(1,0){6}}
\multiput(6,-4)(0,20){2}{\framebox(8,8){\tiny$H$}}
\multiput(14,0)(0,20){2}{\line(1,0){6}}
\multiput(10,7)(0,3){3}{\makebox(0,0){.}}
\end{picture}%
\begin{picture}(62,100)
\multiput(0,0)(0,20){2}{\line(1,0){7}}
\multiput(0,30)(0,25){2}{\line(1,0){7}}
\multiput(0,65)(0,20){2}{\line(1,0){7}}
\multiput(0,95)(0,15){2}{\line(1,0){7}}
\put(7,-8){\framebox(48,126){$\DFT^{-1}$}}
\multiput(55,0)(0,20){2}{\line(1,0){7}}
\multiput(55,30)(0,25){2}{\line(1,0){7}}
\multiput(55,65)(0,20){2}{\line(1,0){7}}
\multiput(55,95)(0,15){2}{\line(1,0){7}}
\end{picture}%
\begin{picture}(30,100)
\multiput(0,0)(0,20){2}{\line(1,0){10}}
\multiput(0,30)(0,25){2}{\line(1,0){10}}
\multiput(0,65)(0,20){2}{\line(1,0){10}}
\multiput(0,95)(0,15){2}{\line(1,0){10}}
\multiput(5,6)(0,4){3}{\makebox(0,0){.}}
\multiput(5,38.5)(0,4){3}{\makebox(0,0){.}}
\multiput(5,71)(0,4){3}{\makebox(0,0){.}}
\multiput(5,100)(0,2.5){3}{\makebox(0,0){.}}
\multiput(15,10)(0,65){2}{\makebox(0,0)[l]{$\left.\rule{0pt}{14\unitlength}\right\}kK$ qubits}}
\put(15,42.5){\makebox(0,0)[l]{$\left.\rule{0pt}{18\unitlength}\right\}k(N-2K-1)$ qubits}}
\put(15,102.5){\makebox(0,0)[l]{$\left.\rule{0pt}{10\unitlength}\right\}k$ qubits}}
\end{picture}%
\kern35pt}\smallskip
\aaeccfigurecaption{Encoder for a quantum Reed--Solomon code.
\label{QRSencoder}}
\end{figure}
In Figure~\ref{QRSencoder} a quantum circuit for encoding quantum RS
codes is presented. The $k(N-2K)$ qubit input state $\ket{\phi}$ is
transformed into a superposition of different cosets of the RS code.
These cosets are determined in the frequency domain, followed by the
quantum version of an inverse Fourier transform $\DFT^{-1}$ over
$\F_{2^k}$. The $\DFT$ is also used for decoding. The syndromes for
bit--flip and phase--flip errors are computed in the frequency domain
(see Figure~\ref{QRSdecoder2}).

\begin{figure}[hbt]
\centerline{\unitlength0.7pt\scriptsize
\begin{picture}(80,160)(-70,0)
\multiput(0,0)(0,30){3}{\line(1,0){10}}
\multiput(0,20)(0,30){3}{\line(1,0){10}}
\put(0,90){\line(1,0){10}}
\put(0,105){\line(1,0){10}}
\multiput(0,115)(0,20){2}{\line(1,0){10}}
\multiput(0,145)(0,10){2}{\line(1,0){10}}
\multiput(5,6)(0,4){3}{\makebox(0,0){.}}
\multiput(5,36)(0,4){3}{\makebox(0,0){.}}
\multiput(5,66)(0,4){3}{\makebox(0,0){.}}
\multiput(5,95)(0,2.5){3}{\makebox(0,0){.}}
\multiput(5,121)(0,4){3}{\makebox(0,0){.}}
\multiput(5,148)(0,2){3}{\makebox(0,0){.}}
\multiput(-1,0) (0,30){2}{\makebox(0,0)[r]{$\ket{0}$}}
\multiput(-1,20)(0,30){2}{\makebox(0,0)[r]{$\ket{0}$}}
\multiput(-12,10)(0,30){2}{\makebox(0,0)[r]{$kK$ qubits
$\left\{\rule{0pt}{16
\unitlength}\right.$}}
\put(-12,107.5){\makebox(0,0)[r]{\begin{tabular}{l@{}}(erroneous)\\encoded\\
state\end{tabular}$\left\{\rule{0pt}{55\unitlength}\right.$}}
\end{picture}%
\begin{picture}(30,100)
\multiput(0,0)(0,30){2}{\line(1,0){30}}
\multiput(0,20)(0,30){2}{\line(1,0){30}}
\multiput(0,60)(0,20){2}{\line(1,0){7}}
\multiput(0,90)(0,15){2}{\line(1,0){7}}
\multiput(0,115)(0,20){2}{\line(1,0){7}}
\multiput(0,145)(0,10){2}{\line(1,0){7}}
\put(7,56){\framebox(16,103){\rotate{$\DFT$}}}
\multiput(23,60)(0,20){2}{\line(1,0){7}}
\multiput(23,90)(0,15){2}{\line(1,0){7}}
\multiput(23,115)(0,20){2}{\line(1,0){7}}
\multiput(23,145)(0,10){2}{\line(1,0){7}}
\end{picture}%
\begin{picture}(30,100)
\multiput(0,0)(0,30){2}{\line(1,0){30}}
\multiput(0,20)(0,30){2}{\line(1,0){30}}
\multiput(0,60)(0,20){2}{\line(1,0){30}}
\multiput(0,90)(0,15){2}{\line(1,0){30}}
\multiput(0,115)(0,20){2}{\line(1,0){30}}
\multiput(0,145)(0,10){2}{\line(1,0){30}}
\multiput(5,135)(20,-20){2}{\line(0,-1){87.5}}
\multiput(5,135)(20,-20){2}{\makebox(0,0){$\bullet$}}
\multiput(5,50)(20,-20){2}{\circle{5}}
\multiput(11,44)(4,-4){3}{\makebox(0,0){.}}
\multiput(11,129)(4,-4){3}{\makebox(0,0){.}}
\end{picture}%
\begin{picture}(20,100)
\multiput(0,0)(0,30){2}{\line(1,0){20}}
\multiput(0,20)(0,30){2}{\line(1,0){20}}
\multiput(0,90)(0,15){2}{\line(1,0){20}}
\multiput(0,115)(0,20){2}{\line(1,0){20}}
\multiput(0,145)(0,10){2}{\line(1,0){20}}
\multiput(0,60)(0,20){2}{\line(1,0){6}}
\multiput(14,60)(0,20){2}{\line(1,0){6}}
\multiput(6,56)(0,20){2}{\framebox(8,8){\tiny$H$}}
\multiput(10,67)(0,3){3}{\makebox(0,0){.}}
\end{picture}%
\begin{picture}(30,100)
\multiput(0,0)(0,30){2}{\line(1,0){30}}
\multiput(0,20)(0,30){2}{\line(1,0){30}}
\multiput(0,60)(0,20){2}{\line(1,0){30}}
\multiput(0,90)(0,15){2}{\line(1,0){30}}
\multiput(0,115)(0,20){2}{\line(1,0){30}}
\multiput(0,145)(0,10){2}{\line(1,0){30}}
\multiput(5,80)(20,-20){2}{\line(0,-1){62.5}}
\multiput(5,80)(20,-20){2}{\makebox(0,0){$\bullet$}}
\multiput(5,20)(20,-20){2}{\circle{5}}
\multiput(11,14)(4,-4){3}{\makebox(0,0){.}}
\multiput(11,74)(4,-4){3}{\makebox(0,0){.}}
\end{picture}%
\begin{picture}(20,100)
\multiput(0,0)(0,30){2}{\line(1,0){20}}
\multiput(0,20)(0,30){2}{\line(1,0){20}}
\multiput(0,90)(0,15){2}{\line(1,0){20}}
\multiput(0,115)(0,20){2}{\line(1,0){20}}
\multiput(0,145)(0,10){2}{\line(1,0){20}}
\multiput(0,60)(0,20){2}{\line(1,0){6}}
\multiput(14,60)(0,20){2}{\line(1,0){6}}
\multiput(6,56)(0,20){2}{\framebox(8,8){\tiny$H$}}
\multiput(10,67)(0,3){3}{\makebox(0,0){.}}
\end{picture}%
\begin{picture}(30,100)
\multiput(0,0)(0,30){2}{\line(1,0){30}}
\multiput(0,20)(0,30){2}{\line(1,0){30}}
\multiput(0,60)(0,20){2}{\line(1,0){7}}
\multiput(0,90)(0,15){2}{\line(1,0){7}}
\multiput(0,115)(0,20){2}{\line(1,0){7}}
\multiput(0,145)(0,10){2}{\line(1,0){7}}
\put(7,56){\framebox(16,103){\rotate{$\DFT^{-1}$}}}
\multiput(23,60)(0,20){2}{\line(1,0){7}}
\multiput(23,90)(0,15){2}{\line(1,0){7}}
\multiput(23,115)(0,20){2}{\line(1,0){7}}
\multiput(23,145)(0,10){2}{\line(1,0){7}}
\end{picture}%
\begin{picture}(90,155)
\multiput(0,0)(0,30){3}{\line(1,0){10}}
\multiput(0,20)(0,30){3}{\line(1,0){10}}
\put(0,90){\line(1,0){10}}
\put(0,105){\line(1,0){10}}
\multiput(0,115)(0,20){2}{\line(1,0){10}}
\multiput(0,145)(0,10){2}{\line(1,0){10}}
\multiput(5,6)(0,4){3}{\makebox(0,0){.}}
\multiput(5,36)(0,4){3}{\makebox(0,0){.}}
\multiput(5,66)(0,4){3}{\makebox(0,0){.}}
\multiput(5,95)(0,2.5){3}{\makebox(0,0){.}}
\multiput(5,121)(0,4){3}{\makebox(0,0){.}}
\multiput(5,148)(0,2){3}{\makebox(0,0){.}}
\put(11,10){\makebox(0,0)[l]{$\left.\rule{0pt}{14\unitlength}\right\}$
\def\arraystretch{0.75}\begin{tabular}{@{}l@{}}syndrome of\\phase--flip\\errors\end{tabular}}}
\put(11,40){\makebox(0,0)[l]{$\left.\rule{0pt}{14\unitlength}\right\}$
\def\arraystretch{0.75}\begin{tabular}{@{}l@{}}syndrome of\\bit--flip\\errors\end{tabular}}}
\multiput(11,70)(0,55){2}{\makebox(0,0)[l]{$\left.\rule{0pt}{14\unitlength}\right\}kK$
  qubits}}
\put(11,97.5){\makebox(0,0)[l]{$\left.\rule{0pt}{10\unitlength}\right\}$
\begin{tabular}{@{}l@{}}$k(N-2K-1)$\\qubits\end{tabular}}}
\put(11,150){\makebox(0,0)[l]{$\left.\rule{0pt}{7\unitlength}\right\}k$ qubits}}
\end{picture}%
}
\aaeccfigurecaption{\parbox[t]{\aaecconecolumn}{Quantum circuit for
computing the syndrome for a quantum Reed--Solomon code.
\label{QRSdecoder2}}}
\end{figure}
\smallskip
\aaeccsection{Conclusion}
\smallskip
Most quantum error--correcting codes known so far are based on
classical binary codes or codes over $GF(4)=\F_{2^2}$
(see~\cite{CRSS98}). We have demonstrated how codes over extension
fields of higher degree can be used. They might prove useful, e.\,g.,
for concatenated coding.

The spectral techniques for encoding and decoding presented do not
only apply to Reed--Solomon codes, but in general to all cyclic
codes. The main advantage of Reed--Solomon codes is that no field
extension is necessary. The same is true for all BCH codes of length
$n$ over the field $\F_{2^k}$ where $n|2^k-1$.  In addition to the
spectral techniques, cyclic codes provide a great variety of
encoding/decoding principles, e.\,g., based on linear shift registers
that can be translated into quantum algorithms (see~\cite{GrBe98}).

The quantum implementation of linear mappings over finite fields
presented in the full paper enlarges the set of efficient quantum
subroutines. In contrast, the transforms used in most quantum
algorithms---such as cyclic and generalized Fourier transforms---are
defined over the complex field (see, e.\,g., \cite{PRB98}).

It has to be investigated how efficient fully quantum algorithms for
error--correction can be obtained, e.\,g., using quantum versions of
the Berlekamp--Massey algorithm or of the Euclidean algorithm.

\smallskip

\aaeccacknowledgements
The authors would like to thank Martin R{\"o}tteler and Rainer
Steinwandt for numerous stimulating discussions during the process of
writing this paper.
\end{aaeccpaper}

\begin{aaeccreferences}
\bibitem{CRSS98}
{\sc A.~R. Calderbank, E.~M. Rains, P.~W. Shor, and N.~J.~A. Sloane}, {\em
  {Quantum Error Correction Via Codes over GF(4)}}, IEEE Transactions on
  Information Theory, \ IT--44 (1998), pp.~1369--1387.

\bibitem{CiZo95}
{\sc J.~I. Cirac and P.~Zoller}, {\em {Quantum Computation with Cold Trapped
  Ions}}, Physical Review Letters, 74 (1995), pp.~4091--4094.

\bibitem{CFH96}
{\sc D.~G. Cory, A.~F. Fahmy, and T.~F. Havel}, {\em {Ensemble Quantum
  Computing by Nuclear Resonance Spectroscopy}}, Tech. Rep. TR--10--96,
  B.\,C.\,M.\,P., Harvard Medical Medical School, Boston, Dec. 1996.

\bibitem{GrBe98}
{\sc M.~Grassl and Th.~Beth}, \ {\em Codierung und Decodierung zykli\-scher
  Quantencodes}, in Fachtagung Informations-- und Mikrosystemtechnik,
  B.~Michaelis and H.~Holub, eds., Magdeburg, 25--27 Mar. 1998.

\bibitem{Gro96_search}
{\sc L.~K. Grover}, {\em {A fast quantum mechanical algorithm for database
  search}}, in Proc. 28th Annual ACM Symposium on Theory of
  Computing (STOC), New York, 1996, ACM, pp.~212--219.

\bibitem{KnLa97}
{\sc E.~Knill and R.~Laflamme}, {\em {Theory of quantum
  error--correcting codes}}, Physical Review~A, 55 (1997), pp.~900--911.

\bibitem{PRB98}
{\sc M.~P{\"u}schel, M.~R{\"o}tteler, and Th.~Beth}, {\em {Fast Quantum Fourier
  Transforms for a Class of Non--abelian Groups}}, in Proceedings
  AAECC--13, 1999.

\bibitem{Shor94}
{\sc P.~W. Shor}, {\em {Polynomial--Time Algorithms for Prime Factorization and
  Discrete Logarithms}}, in Proc. 35th Annual Symposium on
  Foundations of Computer Science (FOCS), 
  IEEE Computer Society Press, Nov. 1994,
  pp.~124--134.

\bibitem{Ste96:multiple}
{\sc A.~Steane}, {\em {Multiple Particle
  Interference and Quantum Error Correction}}, Proceedings of the
  Royal Society  London Series~A, 452 (1996), pp.~2551--2577.

\end{aaeccreferences}
\end{document}